\newcommand{\N}{{N}}
\newcommand{\Z}{{Z}}
\newcommand{\m}{{m}}
\newcommand{\x}{{x}}

\newcommand{\pp}{{p}}
\newcommand{\be}{\begin{equation}}
\newcommand{\ee}{\end{equation}}
\newcommand{\bea}{\begin{eqnarray}}
\newcommand{\eea}{\end{eqnarray}}
\newcommand{\A}{\hat{\mathcal{A}}}
 
 \newcommand{\expect}[2]{\left< \hat{ #1 } \right>_{ #2 }} 
\let\baraccent=\= 
\renewcommand{\=}[1]{\stackrel{#1}{=}} 

\documentclass[prc,aps,preprintnumbers,amsmath,amssymb,showpacs]{revtex4}
\usepackage{graphicx}\usepackage{epsfig}

\begin{document}
\title{Boundary conditions for star matter}
\title{Boundary conditions for fermionic systems within the variational approach }
\title{Boundary conditions for star matter and other periodic fermionic systems}

\author{F. Gulminelli$^{1,2,3}$, T.Furuta$^{1,2,3,4}$,  O.Juillet$^{1,2,3}$, C.Leclercq$^{1,2,3}$} 

\affiliation{
$^{1}$ ENSICAEN, UMR6534, LPC, F-14050 Caen, France \\
$^{2}$ Universit\'e de Caen-Basse Normandie, UMR6534, LPC, F-14032 Caen, France \\
$^{3}$ CNRS, UMR6534, LPC, F-14050 Caen, France \\ 
$^{4}$ present address:  RIKEN Nishina Center, RIKEN, Wako, Japan
}

\begin{abstract}
Bulk fermionic matter, as it can be notably found in supernova matter and neutrons stars, 
is subject to correlations of infinite range due to the antisymmetrisation of the N-body wave function, 
which cannot be 
explicitly accounted for in a practical simulation. This problem is usually addressed in 
condensed matter physics by means of the so-called Twist Averaged Boundary Condition method.
A different ansatz based on the localized Wannier representation has been proposed in the 
context of antisymmetrized molecular dynamics. In this paper we work out the formal relation
between the two approaches. We show that, while the two coincide when working with exact eigenstates
of the N-body Hamiltonian, differences appear in the case of variational approaches,
which are currently used for the description of stellar matter. Some model applications 
with Fermionic Molecular Dynamics are shown.  
\end{abstract}


\today

\maketitle

\section{Introduction}

Interacting fermionic systems in the bulk limit are a standard object of theoretical study 
in condensed matter physics. Electrons are subject to an external periodic potential in the presence of a cristalline ionic structure and the bulk limit can be seen as an infinite number of spatial replicas of a finite system within a specific geometry\cite{book}. 
In that case, the observables of the bulk system can be obtained from the modelization of one single elementary cell, provided adequate boundary conditions are applied to the many-body wave function.
This amounts to introduce a Bloch phase or twist to each wave function in the single-particle basis, 
and average the twisted observables over the different phases within the first Brillouin zone\cite{tabc,kolo}.
In practical applications, this technique has been applied to Quantum Monte-Carlo simulations of electron
systems also in the absence of any external periodic potential\cite{tabc,kolo,mc}.
In this case, the introduction of Bloch phases has to be understood as a technique to accelerate the convergence towards the thermodynamic limit of these very expensive numerical calculations, which would otherwise become prohibitive in computation time. In the absence of an external potential, the periodic cell is just a computation cell with no physical meaning, and independence of the results respect to its size has to be checked.

Coming to the strongly interacting fermionic systems studied in nuclear physics,  the bulk limit has not attracted much interest in the community since nuclei are finite. However there are physical situations   where the propertis of fermionic matter composed of protons and neutrons in the thermodynamic limit are essential such as for core-collapsing supernova, and for the crust of the neutron stars which are left over by the explosion. This nucleonic stellar matter covers a very wide domain of densities ranging from $\rho\approx 10^8$ g $\cdot$ cm$^{-3}$  
to a few times the normal saturation nuclear density $\rho\approx 10^{14}$ g $\cdot$ cm$^{-3}$, temperatures between less than 1 and more than 20 MeV, and proton fractions varying between 0. and 0.5
In the sub-saturation density regime, it is well established that matter is charge neutral and mainly composed of neutrons, protons, electrons, positrons and photons in thermal and typically also chemical equilibrium \cite{prakash_science,haensel_book}. 
Depending on the thermodynamic condition, neutrinos and anti-neutrinos can also participate 
to the equilibrium. 

A very large amount of literature exists on the microscopic modelizations of the neutron star outer and inner crust\cite{San04, Kha08, Gra08,Mon07,constanca,douchin}.
In this regime at zero temperature and relatively low density, matter consists of a lattice of Wigner-Seitz cells, each cell containing a spherical neutron-rich nucleus immersed in a sea of dilute gas of neutrons and relativistic electrons uniformly distributed inside the cell\cite{Neg73, Pet95}. 
The linear size of the cell is of several hundreds of fermis in the outer crust. The size decreases going towards
the center of the neutron star and the central nucleus becomes heavier and increasingly neutron rich.
Typical values\cite{Neg73} in the inner crust range from about 200 particles in a cell of linear size of about 50 fm for $\rho\approx 10^{-4}fm^{-3}$ to about 1500 particles in a cell of linear size of about 15 fm for $\rho\approx 8\cdot 10^{-2}fm^{-3}$. 
Most of the existing calculations are based on variational approaches (Hartree-Fock or Hartree-Fock-Bogoliubov) and employ mixed Dirichlet-Neumann boundary conditions
at the edge of the cell\cite{Neg73}, chosen to produce a flat density close to the cell border and thus to simulate a uniform neutron gas. The specific way of fixing these mixed boundary conditions is not completely clear, and discrepancies due to the choice of fixing these conditions increase with density\cite{Bal06}. A more conceptual problem with Dirichlet or Neumann boundary conditions, is that they neglect antisymmetrization correlations which extend beyond the cell size.
This has been pointed out in refs.\cite{bands,Cha07,Cha09,Has08}, where Bloch boundary conditions, similar to the ones used for the QMC modelizations of bulk electron systems, have been employed. In these works it was shown that both the neutron specific heat and the motion of unbound neutrons are affected by these effects.

In the intermediate region between crust and core, complex phases are predicted that can break the spherical symmetry of the Wigner-Seitz cell, and can violate the associated translational invariance. This is even more true at finite temperature, where the periodicity of the Wigner-Seitz cell is broken by thermal agitation.
These conditions are met in stellar matter in the pre- and post-bounce supernova dynamics, as well as in the cooling process of proto-neutron stars. Even  when  periodicity cannot rigourously be assumed in these thermodynamic conditions, it can be kept  as a practical working hypothesis allowing to address
the thermodynamic limit. As already mentioned, the drawback of that is that convergence with respect to the cell size has to be systematically checked.  
Time-dependent variational microscopic calculations have been proposed to address this region, where statistical averages are calculated from time averages assuming ergodicity\cite{horowitz,watanabe,newton,sebille}. These works have shown that structures, though they can be degenerate in energy\cite{sebille}, are nonetheless approximately periodic in space even at finite temperature. 

In these calculations simple periodic boundary conditions are employed, completely neglecting the antisymmetrization correlations beyond the calculation grid.
The importance of properly accounting for these correlations was recently stressed in ref.\cite{klaas}.
In this work, a specific ansatz for the boundary conditions based on the localized Wannier representation has been proposed in the context of antisymmetrized molecular dynamics. It was shown that such boundary conditions allow obtaining a distribution similar to the one of a free Fermi gas, if gaussian wave packets of fixed width are periodically disposed on a two-dimensional grid, while an artificial Pauli potential is needed in order to obtain the same result with classical molecular dynamics. This result implies that, in the inner crust region where stellar matter contains an important component of quasi-free neutrons, properly accounting for the periodic character of the system may be of importance. 

To conclude these introductory remarks, it appears that simple Dirichlet, Neumann, or periodic boundary conditions are not adapted to the description of bulk fermionic matter. Different solutions are proposed 
in the literature for different applications, namely the Twist Averaged Boundary Conditions (TABC)
in QMC\cite{tabc,kolo}, the Bloch method in mean-field calculations\cite{bands,Cha07, Cha09, Has08}, and the 
Wannier replica method for molecular dynamics approaches\cite{klaas}, but the equivalence of the different techniques and/or their domain of validity is not completely clear.

In this paper, we will formally develop the link between the different methods and show some model applications for simple non-interacting nuclear systems described through the variational Fermionic molecular Dynamic (FMD) method. We will show that Bloch (or TABC) and the replica method are equivalent when they are applied to the exact eigenstates of the many-body Hamiltonian.
When this is not the case, as for variational mean-field theories applied to interacting systems, the equivalence is broken. In this case, the Bloch or TABC technique can produce solutions which differ from the exact results more than if simple periodic boundary conditions are applied. Conversely, the replica method appears more powerful and can easily be applied to any variational based mean-field treatment with an affordable extra computational cost. Our numerical applications will concern one-dimensional systems, for which it has been recently argued\cite{drummond} that additional drawbacks appear in the TABC method. However it is important to remark that one-dimensional modelizations can often be used in the stellar matter case where (except at very high density close to saturation) spherical symmetry is often a good approximation.  

\section{the Bloch theorem and Twist Averaged Boundary Conditions}
 
We want to address the physical problem of an infinite system (specifically: the neutron star crust) 
constituted of an infinite number of spatial replicas of finite systems of linear size $L$ (the Wigner-Seitz cells). In the absence of any periodic external potential, for the replicated system to be equivalent to the infinite system, each Wigner-Seitz cell is supposed to contain an integer number of structures (spherical nuclei or more exotic 'pasta' structures\cite{horowitz,watanabe,newton,sebille}) of size $\vec{l}$,
$L_i=n_i l_i$, $i=x,y,z$. The minimal choice is to take a box containing exactly one structure, and in this case the symmetry of the box will follow the symmety of the physical system (a cylindrical box for cylindrical structures, etc).To simplify the notations we will work in one dimension only and write $L=nl$. The extension to three dimensions is straightforward.

\subsection{The Bloch theorem at the N-body level}
 
We are interested in the translational invariance properties of the Hamiltonian induced by the imposed periodicity.
The Hamiltonian of the global system is obviously invariant respect to a simultaneous translation of all particle coordinates of the same arbitrary length $r$. This invariance physically corresponds to the general statement that the center of mass momentum is a good quantum number. This symmetry has no influence in an infinite system, and we will not consider it further. 
The fact of working in a finite box of linear size $L$ which is replicated induces additionally an extra non trivial invariance, which we now discuss.
The total Hamiltonian of the infinitely replicated system reads
\be
\hat{H}=\sum_{m=1}^\infty {H}_L\left (\hat{\vec{x}}^{(m)}\right ) \label{start}
\ee
where $\vec{x}^{(m)}\equiv (x_1^m,\dots,x_N^m)$ denotes coordinates of particles belonging to the $m$-th replica
and the cell Hamiltonian  is 
\be
\hat{H}_L=\sum_{i=1}^N \hat{t}_i + \sum_{\m=-\infty}^{\infty}\sum_{i>j}^N v\left (\hat{\x}_i-\hat{\x}_j - \m L \right ), \label{h}
\ee
Note that the cell Hamiltonian depends on a finite number $N$ of particles even if these particles are not necessarily confined into the specific volume of the cell. 
This Hamiltonian is invariant under the translation of any particle coordinate $x_k$ of a length $\m L$, with $\m$ integer.
This invariance can be expressed as
\be
\left [ \hat{H}_L,\hat{T}_k(\m) \right ] =0
\ee
where $\hat{T}_k(m)=exp\left[-\frac{i}{\hbar}\m L \cdot \hat{\pp}_k\right]$ is the $L$-translational operator of particle $k$.
Translational invariance implies that the eigenfunctions $\Psi\left (\x_1,\dots,\x_N\right )$ of the cell Hamiltonian eq.(\ref{h}) can be written as
\be
\hat{T}_k(\m) \Psi_{\theta_{k,\m}} = \exp \left (-i\theta_{k,\m}\right ) \Psi_{\theta_{k,\m}}
\label{transl_op}
\ee
where $\theta_{k,\m}$ is the eigenvalue associated to the translation $\x_k \to \x_k-\m L$. Because of the property of translational operators 
$\hat{T}_k(\m_1)\hat{T}_k(\m_2)=\hat{T}_k(\m_1+\m_2)$, the eigenvalues satisfy $\theta_{k,\m_1}+\theta_{k,\m_2}=\theta_{k,\m_1+\m_2}$.
This means that we can associate the translational invariance of periodicity $L$
for particle $k$ with an eigenvalue $\theta_k$ such that 
\be
\theta_{k,\m}=\m \theta_k
\ee

Let us define an auxiliary wave function as
\be
\Phi\left ( \x_1, \dots,\x_N\right )=\exp \left ( -i\frac{\theta_k}{L}\x_k \right )
\Psi\left ( \x_1, \dots,\x_N\right ) \label{Phi}
\ee
Using the fact that $\Psi$ is an eigenfunction of the translation operator eq.(\ref{transl_op}),
%
%
we get
\bea
\Phi\left ( \x_1, \dots,\x_N\right )&=&
\exp \left ( -i\frac{\theta_k}{L}(\x_k -\m L) \right )
   \exp \left ( - i \m \theta_k \right ) \Psi\left ( \x_1, \dots,\x_N\right )\nonumber \\
&=&   
\Phi\left ( \x_1, \dots,\x_k-\m L,\dots,\x_N\right )
\eea
We have shown that, if $\Psi$ is an eigenfunction, then the function $\Phi$ defined by eq.(\ref{Phi}) is a periodic function of period $L$. The same reasoning can be done for any particle $k=1,\dots,N$.
This means that the eigenfunctions of the translationally invariant Hamiltonian eq.(\ref{h}) can be written as
\be
\Psi\left ( \x_1, \dots,\x_N\right )=\exp \left ( i\frac{1}{L}
\sum_{k=1}^{N}\theta_k \x_k \right )
\Phi\left ( \x_1, \dots,\x_N\right )
\label{Psi} 
\ee
where $\Phi$ is a periodic function, that is invariant under the translation $\m L$ of any particle coordinate.

Since the Hamiltonian  eq.(\ref{h}) is invariant under the translation of any particle coordinate separately, one could consider in principle 
a different phase $\theta_k$ for each particle. 
However the indistinguishability of nucleons imposes that all the phases must be equal, as we now show\cite{mc}. 
Let us consider a phase $\theta_1$ for the $L$-translation of particle $1$, and a phase $\theta_2$ for the $L$-translation of particle $2$:
\bea
\Psi\left ( \x_1+L,\x_2, \dots,\x_N\right )&=& \exp\left ( i \theta_1 \right )\Psi\left ( \x_1,\x_2, \dots,\x_N\right ) \label{theta1}\\
\Psi\left ( \x_1,\x_2+L, \dots,\x_N\right )&=& \exp \left ( i \theta_2 \right ) \Psi\left ( \x_1,\x_2, \dots,\x_N\right )
\eea
Applying the permutation symmetry to eq.(\ref{theta1}) gives 
\be
\Psi\left ( \x_1,\x_2, \dots,\x_N\right ) =  - \exp\left (  -i \theta_1 \right ) \Psi\left ( \x_2,\x_1+L, \dots,\x_N\right )
\ee
Applying a translation $-L$ to the second coordinate gives
\be
\Psi\left ( \x_1,\x_2, \dots,\x_N\right ) =  - \exp\left ( -i \left ( \theta_1 - \theta_2 \right )\right ) \Psi\left ( \x_2,\x_1, \dots,\x_N\right )
\ee
Applying the permutation symmetry once again we get
\be
\Psi\left ( \x_1,\x_2, \dots,\x_N\right ) =  \exp\left ( -i \left ( \theta_1 - \theta_2 \right )\right ) \Psi\left ( \x_1,\x_2, \dots,\x_N\right )
\ee
which shows that the two phases must be equal, $\theta_1=\theta_2=\theta_L$.
To be precise we could have $\theta_1-\theta_2=2n\pi$ with any integer $n$,
but  the phase $\theta_L$ can be taken without 
any loss of generality in the interval $(0,2\pi]$ or $(-\pi,\pi]$, that is in the first Brillouin zone. 
Indeed if we consider a very large number $N_r$ of replicas of the WS cell, $N_r \to \infty$, the effect of the Bloch phase is negligible and we can write global periodic boundary conditions for the wave function
\be
\Psi\left ( \x_1+\N_r L, \dots,\x_N+\N_r L\right )=\Psi\left ( \x_1, \dots,\x_N\right )
\label{periodic}
\ee 
%
%
%
This implies $\exp (i\theta_LN_r)=1$ or $\theta_L=(2n-N_r)\pi /N_r$ with $n$ integer.
Let us write $n=n'N_r+n"$ with $1\leq n" \leq N_r$ and $n'$ integer, then
\be
\theta_L=  \left (2n'-1\right ) \pi  + \theta
\ee
with $-\pi<\theta\leq \pi$.
Finally the wave function is
\be
\Psi\left ( \x_1, \dots,\x_N\right )=\exp \left ( i \frac{\theta}{L} \sum_{i=1}^N \x_i \right )\Phi\left ( \x_1, \dots,\x_N\right )
\label{bloch}
\ee
This Bloch formulation is a possible way to represent 
the eigenfunction of the exact N-body Hamiltonian in the cell, accounting for the infinite range antisymmetrization correlations with the replicated system. It amounts to introducing an extra quantum 
number $\theta$, 

\be
\Psi\left ( \x_1, \dots,\x_N\right )=\langle \x_1, \dots,\x_N | \vec{n},\theta \rangle \label{bloch_tot}
\ee

where $\vec{n}$ denotes the ensemble of the other good quantum numbers.
The system observables will thus explicitly depend on the phase $\theta$. 

The form of the periodic Hamiltonian eq.(\ref{start}) implies that the wave function of the global system
can be expressed as an antisymmetrized product of $N$-body Bloch wave functions (\ref{bloch}) according to
\be
\Psi_{tot}\left ( \x_1, \dots,\x_\infty  \right )=\A \prod_{m=1}^{\infty} 
\Psi_{\theta_m}\left ( \x_1^m, \dots,\x_N^m\right ) \label{psitot}
\ee
The Twist Averaged Boundary Condition method consists in assuming that the different angles $\theta_m$ in the product are all different. Then the physical value of any observable $\hat O$ can be computed from eq.(\ref{psitot}) with the extra restriction $\theta_m\neq\theta_n$ giving
\be
\langle\hat O\rangle_n^{tot} = \sum_{m=1}^{\infty}    
\langle \vec{n},\theta_m |\hat O| \vec{n},\theta_m \rangle \label{deco}
\ee
We can see that within this decoherence hypothesis among the different phases, possible non-diagonal terms disappear and the observables per unit cell are obtained as simple averages over phases

 
\be
\langle\hat O\rangle_n^L = \frac{1}{2\pi}\int_ {-\pi}^{\pi}d\theta  
\langle \vec{n},\theta |\hat O| \vec{n},\theta\rangle
\label{TABC}
\ee 

We will explicitly show in the following that the parts of the Fock space corresponding to different phases are indeed disjoint in the case of Slater determinants. For more complex eigenstates, eq.(\ref{TABC}) has to be considered as an approximation which however appears very well verified in practical applications\cite{kolo}. 

\subsection{Application to Slater determinants}

In the following we look for an ansatz for $\Phi\left ( \x_1, \dots,\x_N\right )$ to be introduced as a variational approximation to the full N-body problem. 
In particular mean field approaches, including Antisymmetrized Molecular Dynamics (AMD) 
and Fermionic Molecular Dynamics (FMD), are based on a Slater approximation. This means that the variational
ansatz can be written as 
\be
\Phi\left ( \x_1, \dots,\x_N\right )=\A \prod_{k=1}^{N} u_k(\x_k)
\label{slater}
\ee
where $\A$ is the antisymmetrization operator among the N particles and  $u_k(\x+\m L)=u_k(\x)$ $\forall k$.
To fulfill this periodicity condition we can write
\be
u_k(\x)={\cal N} \sum_{m=-\infty}^{\infty}g_k(\x-\m L)
\label{un}
\ee
where $g$ is an arbitrary function and ${\cal N}$ is a normalization factor. 
In the following we will show applications with the non-orthogonal 
single-particle FMD/AMD basis set\cite{fmd,amd} given by non-normalized gaussian wave packets
\be
g_{Z_k}(x)\equiv \exp \left ( - \frac{1}{2a_k}\left (Z_k-x\right)^2\right ) \label{gaussian}
\ee
where $Z_k$, $a_k$ are complex variational parameters.
For this specific choice the translation of the argument is equivalent to a translation of the gaussian centroid $Z_k$
\be
u_k(\x)=\lim_{N_r\to\infty}
\frac{1}{\sqrt{N_r}}\sum_{m=-N_r/2}^{N_r/2}g_{Z_k+\m L}(\x) \label{ansatz1}
\ee
where the dependence on $a_k$ is implicit and omitted to simplify the notations.
It is important to remark that the choice of the normalization in eq.(\ref{ansatz1}) guarantees that the norm
of the wave function is finite, which will allow the numerical evaluations below. In the special case where the gaussian width is sufficiently small respect to the cell size, such that the overlaps between gaussians can be neglected this norm is readily evaluated as $<u_k|u_k>=\sqrt{\pi|a_k|}$. 

Let us consider a generic one-body operator $\hat{O}=\sum_{k=1}^{N}\hat{o}_k$. The matrix element in r-space representation reads
\be
o_{jk}=<u_j|\hat{o}|u_k>=\int_{-\infty}^{\infty}d\x \int_{-\infty}^{\infty}d\x'
u_j^*(\x') o(\x',\x) u_k(\x)
=\lim_{N_r\to\infty}\int_{-\frac{N_r}{2}L}^{\frac{N_r}{2}L}d\x \int_{-\infty}^{\infty}d\x'
u_j^*(\x') o(\x',\x) u_k(\x)
\ee
Because of the periodicity of the $u_k$ this simplifies to
\bea
o_{jk}
&=& \lim_{N_r\to\infty} N_r \int_{-L/2}^{L/2} d\x \int_{-\infty}^{\infty} d\x' u_j^*(\x') o(\x',\x) u_k(\x)
\eea
where we have used the fact that the operator is translationally invariant.
For a local one-body operator $\hat{A}=\sum_{k=1}^{N}\hat{a}_k$ the matrix element simplifies to
\be
a_{jk} =
\lim_{N_r\to\infty} N_r \int_{-L/2}^{L/2} d\x  \  {u}_j^*(\x) a(\x)  {u}_k(\x)
\ee
%
%
%
This shows that, because of the periodicity, only integrals over one single box are needed. 
The expectation value of the observable $\hat{O}$  is given by (for the moment we are ignoring the Bloch phase, and considering expectations only over the auxiliary function $\Phi$, which we indicate $<>_\Phi$):
\be
<\hat{O} >_\Phi = <\Phi|\hat{O}|\Phi> = \sum_{k=1}^{N} o_{kk} \label{phiav1}
\ee
if the $u_k$ constitute an orthonormal basis, and
\be
<\hat{O} >_\Phi = \frac{<\Phi|\hat{O}|\Phi>}{<\Phi|\Phi>} = \sum_{j,k=1}^{N} o_{jk}B_{kj}^{-1} \label{phiav2}
\ee
if not. Here $B_{jk}$ is the overlap matrix
\be
B_{jk}=<u_j|u_k>.
\ee
Implementing the ansatz (\ref{ansatz1}), the matrix element of a generic local operator is calculated by
\be
a_{jk} =
\lim_{N_r,N'_r,N''_r\to\infty} \frac{N''_r}{\sqrt{N_rN'_r}}
\sum_{m=-N_r/2}^{N_r/2}\sum_{m'=-N'_r/2}^{N'_r/2} \int_{-L/2}^{L/2} d\x 
g^*_{\Z_j+\m L}(\x) a(\x) g_{\Z_k+\m' L}(\x) \label{aajk}
\ee
Because of the finite norm of $u_k$, the double sum is a convergent quantity with increasing number of replicas
\bea
&\lim_{N_r,N'_r\to\infty}& \sum_{m=-N_r/2}^{N_r/2}\sum_{m'=-N'_r/2}^{N'_r/2} \int_{-L/2}^{L/2} d\x 
g^*_{\Z_j+\m L}(\x) a(\x) g_{\Z_k+\m' L}(\x) = \\ 
&\lim_{N_r,N'_r\to\infty}&  \frac{1}{\sqrt{N_rN'_r}}\sum_{m=-N_r/2}^{N_r/2}\sum_{m'=-N'_r/2}^{N'_r/2} \int_{-\infty}^{\infty} d\x 
g^*_{\Z_j+\m L}(\x) a(\x) g_{\Z_k+\m' L}(\x) \\ && = <u_k |\hat a|u_k >
\eea

This means  that the limits in eq.(\ref{aajk}) can be performed separately giving:
\be
a_{jk} 
= \int_{-L/2}^{L/2} d\x  \sum_{m,m'=-\infty}^{\infty} 
g^*_{\Z_j+\m L}(\x) a(\x) g_{\Z_k+\m' L}(\x)   
\ee
In the hypothesis that the spatial extension of the wave function is negligible beyond a linear size $L_{max}= M L$,  $\lim_{x\to \pm {L}_{max}}\langle{x}|g_{Z}\rangle=0$,  we can write 
\be
a_{jk}
= \int_{-L/2}^{L/2} d\x  \sum_{m,m'=-M}^{M} 
g^*_{\Z_j+\m L}(\x) a(\x) g_{\Z_k+\m' L}(\x) 
\label{ajk}  
\ee
which is a feasible integral.  
Eq.(\ref{ajk}) is readily generalized to the case of a non-local operator.

Because of the translational invariance the wave function has to be multiplied by the Bloch phase factor eq.(\ref{bloch}). This means that the expectation values of observables have to be corrected respect to eq.(\ref{phiav1}),(\ref{phiav2})
\be
<\hat{O}>_\Psi = 
\sum_{j,k=1}^N o_{jk\theta} B^{-1}_{kj\theta}
= \frac{< {\Psi}_{\theta}|\hat{O}| {\Psi}_{\theta} >}{< {\Psi}_{\theta}| {\Psi}_{\theta} >}
\ee
where the notation $<>_\Psi$ corresponds to a specific choice for the (arbitrary) Bloch phase $-\pi < \theta \leq \pi$,
\be
 {\Psi}_{\theta}(\x_1,\dots,\x_N)=\A 
\prod_{k=1}^N {\psi}_{k\theta}(\x_k),
\ee
\be
 {\psi}_{k\theta}(\x)=\exp\left ({i\theta\x/L}\right )  {u}_{k}(\x)
\ee
We can finally write the expression of the different matrix elements for the infinite periodic system, for a given choice of the Bloch phase $\theta$.
The overlap matrix is not influenced by the Bloch phase, and the same is true for the matrix element of any local operator:
%
%

%
\be
a_{jk\theta}= \sum_{m,m'=-M}^{M} \int_{-L/2}^{L/2} d\x  
g^*_{Z_k+\m L}(\x) a(\x)  g_{Z_k+\m' L}(\x)  = 
a_{jk}
\ee
In particular the local density
\bea
\rho_{\theta}(\x) &=& \sum_{j,k=1}^A <\x| {\psi}_{k}>< {\psi}_{j}|\x>
B_{kj}^{-1} \nonumber \\ 
&=&  \sum_{j,k=1}^A <\x| {u}_{k}>< {u}_{j}|\x>
B_{kj}^{-1} \nonumber = \rho(\x)
\eea
is independent of $\theta$.

The situation is different for non-local operators, because in this case the Bloch phase does not cancel any more
\be
o_{jk\theta}= \sum_{m=-\infty}^\infty \sum_{m'=-M}^{M} \int_{-L/2}^{L/2} d\x
\int_{-\infty}^{\infty} d\x'  
g^*_{Z_j+\m L}(\x') o(\x,\x')  g_{Z_k+\m' L}(\x) 
\exp\left (i\theta(\x-\x')/L\right ) 
\ee
Let us take the exemple of the momentum operator
\bea
\hat{k} {\psi}_{j}(\x) 
&=& -i  \sum_{m=-M}^{M}  
\exp \left ( i \theta x/L \right ) 
\left ( \frac{\partial}{\partial x} +\frac{i\theta}{L} \right )
 g_{Z_j+\m L}(\x) 
\eea
The expectation value of the total momentum in the cell $\hat{K}=\sum_{j=1}^N \hat{k}_j$ 
is given by
\bea
<\hat{K}>_\Psi &=& \sum_{ji=1}^{N} 
<{\psi}_{j}|\hat{k}|{\psi}_{i}> B_{ij}^{-1}
\nonumber \\
&=& \sum_{ji=1}^{N} 
<{u}_{j}|\hat{k}|{u}_{i}> B_{ij}^{-1}
+N\frac{\theta}{L} 
= <\hat{K}>_\Phi +N \frac{\theta}{L} \label{totalP}
\eea
We can see that the Bloch phase physically represents a boost to the center of mass motion. It will cancel when performing the average between the different phases $-\pi < \theta \leq \pi$ (TABC), but this will give a finite contribution to the kinetic energy. Indeed if we consider the square momentum
\be
\hat{k}^2{\psi}_{j}(\x)  
= -  \exp \left ( i \theta x/L \right ) 
\sum_{m=-M}^{M}  
\left ( \frac{\partial^2}{\partial x^2} +\frac{2i\theta}{L}\frac{\partial}{\partial x} 
-\frac{\theta^2}{L^2}
\right )
 g_{Z_j+\m L}(\x) 
\ee
This gives rise to a total kinetic energy expectation value
\bea
<\hat{E}_K>_\Psi &=& \frac{\hbar^2}{2m} \sum_{jk=1}^{N} 
<{\psi}_{j}|\hat{k}^2|{\psi}_{k}> B_{kj}^{-1}
+ \nonumber \\
&+& \frac{\hbar^2}{2m} \frac{2\theta}{L}<\hat{K}>_\Phi + \frac{\hbar^2}{2m} N \frac{\theta^2}{L^2} \nonumber \\
&=& <\hat{E}_K>_\Phi + \frac{\hbar^2\theta}{mL}<\hat{K}>_\Phi + \frac{N}{2m} \left ( \frac{\hbar^2\theta}{L}\right )^2 
\eea
This extra kinetic energy term linked to the Bloch phase explicitly enters in the energy variation (or in the equations of motion, in the case of dynamical models\cite{fmd,amd,sebille,newton}).  
If we consider the typical size of a Wigner-Seitz cell as calculated by Negele and Vautherin\cite{Neg73}, in the inner crust ($L\approx 15$ fm) the phase effect gives rise to an energy contribution $\Delta E\approx 11$ MeV, which is far from being negligible if we compare to the corresponding Fermi energy $E_{F}\approx 24$ MeV
\cite{bands,Cha07,Cha09}.

\subsection{One-dimensional model cases}

As a first model case, we consider one-dimensional free particles. Even if this system is clearly very far from the correlated dishomogeneous solutions relevant for stellar matter, it has the advantage of being exactly solvable. Moreover, it is an especially interesting test case for molecular dynamics models~\cite{fmd,amd,klaas} and more general models employing localized wave functions\cite{sebille}: these models are optimized to treat density fluctuations but not necessarily adapted to treat the coupling to the continuum which is needed for the extreme neutron-proton ratio which is found in the inner crust of neutron stars.  
   
In one dimension, the energy per particle of a degenerate ideal Fermi gas at density
$\overline{\rho}=N/L=\frac{k_F}{\pi}$ is given by $e_{FG}=\frac{\hbar^2\pi^2\overline{\rho}^2}{6m}$.  
In the case of a finite system of $N$ particles within a length $L$ with periodic boundary conditions, 
the energy levels are given by
\be
E_i=\frac{\hbar^2}{2m}\left(\frac{2\pi n_{i}}{L}\right)^2,
\ee
 where $n_{i}$ is an integer within the interval $[-\frac{N}{2},\frac{N}{2}]$. 
  Due to this level degeneracy, the total cell momentum
   $ \langle{\hat K}\rangle^L$ is equal to $\pm \frac{\pi N}{L}$ for an even 
   number of particles, while it is zero for an odd number.   
   
Because of the periodicity of the system, it is easier to work in Fourier space.
All one-body observables can be expressed as a function of the one-body density
in momentum space
 $\rho_{\Psi}(k,k')=\sum_{i,j=1}^{N}\langle{k}|{\psi}_i\rangle 
 \langle {\psi}_j|{k'}\rangle B^{-1}_{ij}$, 
 where the single-particle  states are given in the momentum representation. These states can be obtained
 by taking the Fourier transform of the wave function. Considering that a periodic function can always be expressed as a Fourier series,
\be
u_n(x)=\sum_{l=-\infty}^\infty c_{nl} e^{\left ( - 2\pi i l x/L \right )}
\ee
we get 
\bea
 {\psi}_{n\theta}(k)
&=& {\sqrt{2\pi}}
\sum_{l=-\infty}^\infty c_{nl} \;
\delta\left ( \theta/L - k - 2\pi l/L \right ) \nonumber \\
&=&\lim_{N_r\to\infty} \sum_{l=-\infty}^\infty 
\sum_{m=-\frac{N_r-1}{2}}^{\frac{N_r-1}{2}}
\sqrt{\frac{2\pi}{N_r}}\frac{1}{L}
\int_{-\frac{L}{2}}^{\frac{L}{2}}dx \,
g_{Z_n+mL}(x)e^{2i\pi l x/L} \delta\left ( \theta/L - k - 2\pi l/L \right ) \nonumber \\
&=&
\lim_{N_r\to\infty} \sum_{l=-\infty}^\infty 
\sqrt{\frac{2\pi}{N_r}}\frac{1}{L}
\int_{-\frac{N_r L}{2}}^{\frac{N_r L}{2}}dx \,
g_{Z_n}(x)e^{2i\pi l x/L} \delta\left ( \theta/L - k - 2\pi l/L \right ) \nonumber \\
&=&
\frac{2\pi}{L}\sqrt{\frac{a_n}{N_r}} \sum_{l=-\infty}^\infty 
e^{-\frac{a_n}{2} \left( \frac{2\pi l}{L} \right)^2}e^{i\frac{2\pi l}{L} Z_n}
\delta\left ( \theta/L - k - 2\pi l/L \right ) 
\eea

 


 
At variance with the spatial density, the momentum density thus explicitly depends on the
value of the phase $\theta=L k_0$:
\be 
\rho_\theta(k)=\lim_{N_r\to\infty} 
\frac{4\pi^2}{N_r L^2}
\sum_{n,p=1}^{N} B^{-1}_{n,p}\sqrt{a_n a_p^*}
e^{-\frac{a_n+a_p^*}{2} \left( \frac{2\pi l}{L} \right)^2}e^{i\frac{2\pi l}{L}\left( Z_n -Z_p^* \right)}
\delta\left ( \theta/L - k - 2\pi l/L \right )
\ee
The momentum density  is obtained by averaging over the different phase values, 
giving
\be 
\rho_\Psi(k)=
\lim_{N_r\to\infty} 
\frac{2\pi}{N_r L}
\sum_{n=-\frac{N_r-1}{2}}^{\frac{N_r}{2}} 
\sum_{n,p=1}^{N} B^{-1}_{n,p}\sqrt{a_n a_p^*}
e^{-\frac{a_n+a_p^*}{2} \left( \frac{2\pi l}{L} \right)^2}e^{i\frac{2\pi l}{L}\left( Z_n -Z_p^* \right)}
  \label{rhok}  
\ee

We can see that the value of $\rho$ in $k$ is solely determined by the wave function $\Psi$ corresponding to the phase value $\theta=k-\frac{2\pi}m$, which corresponds to a unique value in the first Brillouin zone. This means that the number of points chosen for the discretization
of the integral over the Bloch phase defines the resolution of the momentum density: the Bloch phase 
generates new plane waves which 'fill up' the distribution as needed to obtain a continuous Fermi sea starting from the discrete levels of a finite system. 

\begin{figure}[htbp]
{
\begin{center}
{
\scalebox{1}{\input{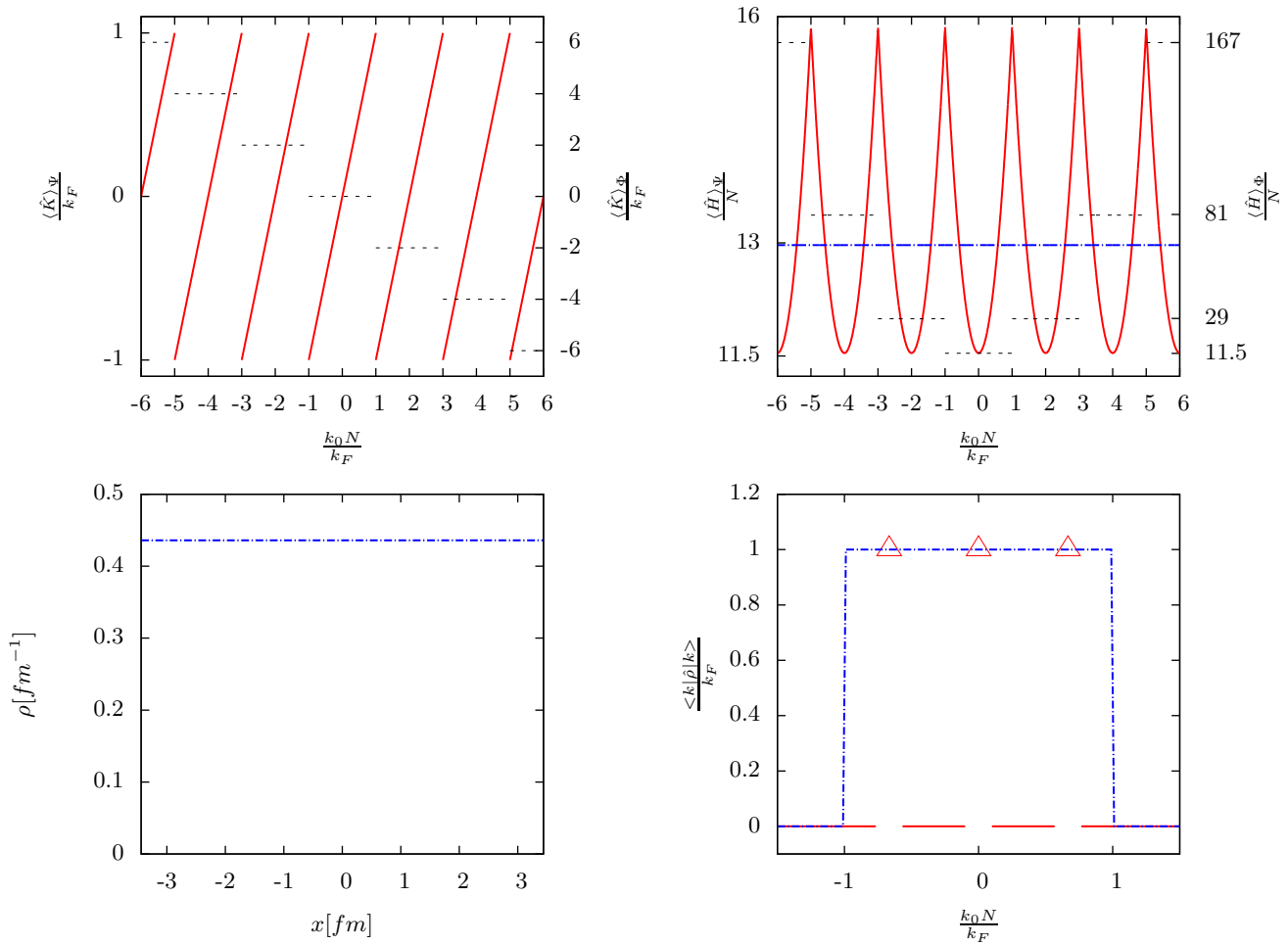}}
}		 
\end{center}
\caption{
FMD calculation of a system of three free particles in a one dimensional box of size $L$ with Bloch boundary conditions. Upper part:
behavior of the total linear momentum (left) and total energy (right)  as a function of the Bloch wave number $k_0=\theta/L$ 
Full lines: total expectation values $\langle\hat K\rangle_\Psi,\langle\hat H\rangle_\Psi$ (see text);
dashed lines: averages over the variational part of the wave function  $\langle\hat K\rangle_\Phi,\langle\hat H\rangle_\Phi$ without the contribution of the Bloch phase (see text);
dash-dotted line: total energy average over the Bloch phases.
Lower part: spatial (left) and momentum (right) density.
}
\label{free}
}
\end{figure}

The total energy and momentum easily follow
\be  
\expect{E}{\Psi}=\frac{\hbar^2}{2m}\int_{-\infty}^{\infty}dk\rho_\Psi(k)k^2, \, \,\, \,
\expect{K}{\Psi}=\int_{-\infty}^{\infty}dk\rho_\Psi(k)k.
\ee
The result for the FMD model is shown in Fig.\ref{free}.
We have chosen a particle density $\overline{\rho}=\frac{1.37}{\pi}$. The size $L$  of the elementary cell is directly proportional to the number of particles in the cell according to
 $L=\frac{N}{\overline{\rho}}$. 
 The FMD ground state is obtained minimizing the total energy with respect to the variational parameters $\vec{z}=\{Z_n,a_n, n=1, \dots, N\}$. This can be obtained using the gradient method, or alternatively by the numerical solution of the FMD equations of motion  \cite{fmd}  
\bea
\frac{d\langle\hat H\rangle_\Psi}{dz_i} &=& \sum_j C^*_{ij} \frac{dz^*_i}{dt}, \label{eom}\\
C_{ij}&=&i\hbar \frac{\partial^2}{\partial z^*_i \partial z_j} \ln \langle \Psi | \Psi\rangle
\eea
It is possible to show\cite{fmd} that if a friction term is  added as a multiplicative factor on the r.h.s. of eq.(\ref{eom})
\be
\frac{d\langle\hat H\rangle_\Psi}{dz_i} = A^* \sum_j C^*_{ij} \frac{dz^*_i}{dt},
\ee
it gives a dynamics of the variational parameters leading to a decrease in time of the total energy.
The resulting dissipative dynamics can be written as:
\be
\frac{d\langle\hat H\rangle_\Psi}{dt} = 2i \Im (A) \sum_{ij} C_{ij} \frac{dz^*_i}{dt}.\frac{dz_j}{dt}
\ee
and this equation is numerically followed until convergence.

As expected, Figure \ref{free} shows that the periodicity in $k$-space is reproduced by the calculation. 
This is due to the implicit  dependence of the expection values $\langle\hat P\rangle_\Phi,\langle\hat H\rangle_\Phi$ on the choice of the Bloch quantum number $k_0=\theta/L$ obtained through the variational procedure.

The exact results of a free Fermi gas are recovered in the model. 
This means that the FMD ansatz for the wave function appears adapted to describe the free particle system. 
This was not a-priori obvious because the choice of single-particle wave packets represents a restriction of the complete Slater determinant variational space, and the plane wave is obtained only as a mathematical limit $a\to\infty$ of the wave function, which is not exactly accessible in numerical simulations. This implies that the FMD model, which has been introduced essentially to describe cluster degrees of freedom, can also address continuum states. 

The other interesting point is that the exact Fermi gas result is obtained with any 
arbitrary number of particles in the simulation. For the 3-particles system shown in Figure \ref{free}, the triangles in the lower right part of the figure show the momentum values allowed for this finite system.
These values are eigenvalues of the periodic part of the wave function $\Phi$, and their spacing decreases with increasing number of particles. As shown by eq.(\ref{rhok}), the effect of the $\theta$ quantum number is to produce all the other missing values for the $k$ quantum number which would be accessible to the infinite system, thus reproducing the continuous Fermi distribution. 
 
In conclusion, the application of TABC to Bloch single particle wave functions appears as a very powerful
method to recover the correct kinetic energies and wave functions at the thermodynamic limit with variational
methods applied to small systems even in one dimension. A word of caution is however necessary: to 
demonstrate the Bloch theorem we have explicitly used the fact that the states are eigenvectors of the many-body Hamiltonian, which is in general not true when using variational methods. 
For this reason we will develop an alternative scheme in the next section.

\section{The replica method}

The twist averaged boundary condition are currently applied in QMC calculations of correlated electron systems at the thermodynamic limit. In the specific framework of fermionic molecular dynamics, an alternative method to deal with the infinite range antisymmetrisation correlations has been proposed, based on the localized Wannier representation\cite{klaas}. Taking advantage of the nested structure of the overlap matrix in a periodic system, an analytical method of inversion of this matrix is proposed in ref.\cite{klaas}. We present here a slightly different derivation of the same equations, for an application to any arbitrary single particle basis. 
 
\subsection{General formulation}

Let us consider that our WS cell contains already many different replicas of the physical system, $N_{tot}=N_r N$ where finally we will let $N_r\to \infty$.
Similarly, we now interpret the simulation cell $L_{tot}=nL$ as a huge length, $n \gg 1$, such that working with a finite $L_{tot}$ can be considered as equivalent to the thermodynamic limit, meaning that  the scaling has to be fulfilled for all observables $\hat{O}$
\be
<\hat{O} > =N_r <\hat{O} >^{L}
\ee
with
\be
<\hat{O} >^{L}=\frac{1}{N_r} \sum_{j,k=1}^{N_{tot}} o_{jk}B^{-1}_{kj}
\label{ophibox}
\ee
independent of the periodicity properties of the Hamiltonian. 
If we consider the $N_{tot}$-body wave function, the invariance property of the system under the translation $\m L_{tot}$ of any particle coordinate is still verified. The difference respect to the previous section is that the system composed of 
$N_{tot}$ particles has an additional symmetry which is not directly linked with the use of a finite box $L$ but is rather due to the physical periodicity of the pasta structures over a lenght scale $L=L_{tot}/n$. 
We have already observed that, because of the absence of an external field, the cell Hamiltonian eq.(\ref{h}), which now represents the full Hamiltonian eq.(\ref{start}),
\be
\hat{H}=\sum_{i=1}^{N_{tot}} \hat{t}_i + \sum_{\m=-\infty}^{\infty}\sum_{i>j}^{N_{tot}} v\left (\hat{\x}_i-\hat{\x}_j - \m L_{tot} \right )
\ee
is invariant under the simultaneous translation of every particle coordinate of any arbitrary length. The translational invariance which will be relevant to us is the 
invariance under the simultaneous translation of every particle coordinate of the physical length $L$, defined as the periodic length of the one body density
\be
\rho(\x+\m L)=\rho(\x) \label{density}
\ee
If the Hartree-Fock field as an external field, this translational invariance would be the only one verified by the mean-field hamiltonian,
%
%
and it would be verified for each particle coordinate separately. The situation is different in the case of an interacting system as described by the self-consistent Hartree-Fock approach. In this case, translational invariance with respect to any arbirary length is  respected because of the self-consistency of the mean-field approach, $U_{ij}=\delta E_{HF} / \delta \rho_{ji}$ where $E_{HF}$ is the interaction part of the Hartree-Fock energy. However, this translational invariance applies only to the simultaneous translation of all particle coordinates. The only special feature of the length $L$ is that eq.(\ref{density})
is fulfilled, which will allow us to impose a simplified expression for the one-body wave functions, as we will see in a moment.

Following the derivation of the previous section, these two invariance properties 
lead to the definition of a Bloch phase for the $N_{tot}$-body wave function according to
\be
\Psi\left ( \x_1, \dots,\x_{N_{tot}}\right )=\exp \left ( i\frac{\theta_l}{{L N_{tot}}}
\sum_{k=1}^{N_{tot}} \x_k \right )
\Phi\left ( \x_1, \dots,\x_{N_{tot}}\right ) 
\ee
where $\Phi$ is a periodic function, that is invariant both under the translation $\m L$ of all ${N_{tot}}$ particle coordinates simultaneously, and under the translation $\m L_{tot}$ of any particle coordinate.

The difference respect to the derivation of the previous section is that now we have a factor $1/{N_{tot}}$ multiplying the Bloch phase.  Since we are at the thermodynamic limit ${N_{tot}}\to\infty$, the correction
on kinetic observables induced by this Bloch phase can be neglected and we can consider that the total $N_{tot}$-body wave function $\Psi$ shares the periodicity properties of $\Phi$. 
 
Let us now look for an ansatz for this $L-$ and $L_{tot}-$periodic $\Psi\left ( \x_1, \dots,\x_{N_{tot}}\right )$ to be introduced as a variational
approximation to the full $N_{tot}$-body problem. 
We take the same Slater ansatz used in the previous section: 
\be
\Psi\left ( \x_1, \dots,\x_{N_{tot}}\right )=\hat {A} \prod_{k=1}^{N_{tot}} u_k(\x)
\label{slater_tot}
\ee
with $u_k(\x+mL_{tot})=u_k(\x)$.
Because of the periodicity of the one body density, each sub-cell of linear dimension $L$ can be exactly associated to a finite number $N=N_{tot}/N_r$ particles, and the wave function can be written introducing a sub-cell index $m$ as: 
\be
\Psi\left ( \x_1, \dots,\x_{N_{tot}}\right )=
\hat{A} \prod_{k=1}^{N} \prod_{m=1}^{N_r} u_{k,m}(\x)
\ee
Moreover, the periodicity of $\rho$ can be written as
\bea
\rho(\x+ L)&=&\sum_{m,m'=1}^{N_r}\sum_{k,j=1}^{N}
u^*_{k,m}(\x+ L) u_{j,m'}(\x+ L)B_{jk}^{-1} \nonumber \\
&=& \sum_{m,m'=1}^{N_r}\sum_{k,j=1}^{N}
u^*_{k,m}(\x) u_{j,m'}(\x)B_{jk}^{-1} = \rho(\x)
\eea
This condition can be satisfied introducing only $N$ independent one body wave functions
$u_{k,m}(\x)=u_{k}(\x-\m L)$
%
%
%
%
where in principle $u_{k}$ are arbitrary (non periodic) functions.
If these functions would form an orthonormal basis
\be
<j,m|k,m'> = \delta_{jk} \delta_{mm'}
\ee
the problem would be reduced to a $N$ body problem as with the ansatz eq.(\ref{slater}). Indeed in the calculation (\ref{ophibox}) of an arbitrary one-body observable only the $u_k$, $k=1,\dots,N$ functions would be needed
\be
<\hat{O} >^{L}=\frac{1}{N_r} \sum_{k=1}^{N_{tot}} o_{kk}
=\sum_{k=1}^N o_{kk}.
\ee

\subsection{Application to non-orthogonal single-particle basis}

For practical applications we will not work with an orthonormal basis, but we want to  take the FMD/AMD gaussian wave packet variational ansatz eq.(\ref{gaussian}), namely
\be
u_{k}(\x-\m l)=g_{\Z_k+\m l} (\x)
\ee
Since the $g_{\Z_k+\m l}$ do not form an orthonormal basis, we are left with a problem of computational size $N_{tot}^2$, that is impossible to solve (recall
$N_{tot}\to\infty$). 

A possibility would be again to impose the simplified expression:
\be
\Psi\left ( \x_1, \dots,\x_{N_{tot}}\right )=\prod_{m=1}^{N_r} \A \prod_{k=1}^{N} g_{\Z_k+\m l}(\x)
=\prod_{m=1}^{N_r} \Psi\left ( \x_1, \dots,\x_N\right )
\label{choice1}
\ee
which allows to work with only $N$ functions.

The conceptual problem with this simple ansatz eq.(\ref{choice1}) is that it neglects  the antisymmetrization correlations among different cells. We know that such correlations are important as they are at the origin of the band structure.
For this reason we need to keep the antisymmetrization correlation among all the $N_{tot}=N_r\cdot N$ particles.

By introducing a linear transformation
\be
\vec{u_k}=\hat{U}\vec{g_k}
\label{uu}
\ee
defined by
\be
u_{k,n}(\x)=\frac{1}{\sqrt{N_r}}\sum_{m=-N_r/2}^{N_r/2} \exp \left ( im\frac{2\pi n}{N_r} \right ) g_{\Z_k+\m L}(\x), \label{unitary}
\ee
%
%
%
we can reduce the computation size to $N_r N^2$ keeping the antisymmetrization correlations among all the $N_{tot}=N N_r$ particles.
It is interesting to remark that, in the limit of $N_r\to\infty$, this transformation coincides with the definition of Wannier functions\cite{book}.
This transformation preserves the determinant
\be
 \hat{A} \prod_{m=1}^{N_r} g_{\Z_k+\m L}(\x) =  \hat{A} \prod_{n=1}^{N_r} u_{k,n}(\x)
\ee
so that we can write
\be
\Psi\left ( \x_1, \dots,\x_{N_{tot}}\right )=\A \prod_{n=1}^{N_r} \prod_{k=1}^{N}  u_{k,n}(\x)
\ee
The advantage of the transformation (\ref{unitary}) is that matrices are block-diagonal, that is functions corresponding to different phases $\gamma=(2n-N_r)\pi/N_r$ are orthogonal
\be
<{u}_{j,n}|{u}_{k,n'}> =   \delta_{n,n'}B_{jk}(n).
\ee
%
We have thus shown that the decoherence hypothesis eq.(\ref{deco}) of the TABC method is exactly verified in the case of variational approaches based on Slater determinants.

Having recovered the orthogonality on the level of the replica quantum number we can write
\be
<\hat{O} >_\Psi^{L}=\frac{1}{N_r} \sum_{n=1}^{N_r} 
\sum_{j,k=1}^N <{u}_{j,n}|\hat{o}|{u}_{k,n}> B_{kj}^{-1}(n)
\ee
The matrix element of any  operator is given by
\be
o_{jk}(n)=<{u}_{j,n}|\hat{o}|{u}_{k,n}>
= \int_{-L/2}^{L/2} d\x  \sum_{m,m'=-\infty}^{\infty} 
g^*_{\Z_j+\m L}(\x') o(\x,\x') g_{\Z_k+\m' L}(\x) \exp\left ( -i (m-m') \frac{2\pi n}{L N_r}\right )
\ee
and explicitly depends on the $n$ quantum number associated to the antisymmetrization among the different subcells. 
In particular the overlap matrix reads
\be
B_{jk}(n)=<{u}_{j,n}|{u}_{k,n}>
= \int_{-L/2}^{L/2} d\x  \sum_{m,m'=-\infty}^{\infty} 
g^*_{\Z_j+\m L}(\x) g_{\Z_k+\m' L}(\x) \exp\left ( -i (m-m') \frac{2\pi n}{L N_r}\right )
\ee
Using the same argument that we had for the expectation (\ref{ajk}), the series can be reduced to a finite sum 
\be
B_{jk}(n)
 = \sum_{m,m'=-M}^{M} \int_{-L/2}^{L/2} d\x  
g^*_{\Z_j+\m L}(\x)   g_{\Z_k+\m' L}(\x)  \exp\left ( -i (m-m') \frac{2\pi n}{L N_r}\right )
\ee
In practical application, a large value of $M\approx 10$ is needed to describe non-interacting particles, but $M=1$ turns out to be sufficient in the presence of the nuclear and Coulomb interaction for the particle densities relevant for the neutron star crust.
From the computational viewpoint, $M=1$ is equivalent to calculate a single gaussian overlap between each pair  if periodic boundary conditions are applied to the gaussian, meaning that the simulation of the infinite system is 
not more expensive numerically than the one for the finite system.
The analogous expression for the matrix elements reads
\be
<{u}_{j,n}|\hat{a}|{u}_{k,n}>
= \sum_{m,m'=-M}^{M} \int_{-L/2}^{L/2} d\x  
g^*_{\Z_j+\m L}(\x) a(\x)  g_{\Z_k+\m' L}(\x)  \exp\left ( -i (m-m') \frac{2\pi n}{L N_r}\right ) 
\ee
Going to the $N_r\to\infty$ limit we can write
\be
\lim_{N_r\to\infty} \frac{2\pi n}{L N_r}=\gamma
\label{gamma}
\ee
with $-\pi <\gamma\leq \pi$.
The observables read
\be
<\hat{O}>_\Psi^{L} =\frac{1}{2\pi} \int_{-\pi}^{\pi} d\gamma \sum_{j,k=1}^N
<{u}_{j,\gamma}|\hat{o}|{u}_{k,\gamma}>
B_{kj}^{-1}(\gamma)
\ee
which is expected to give equivalent results to the implementation where TABC are applied.  

\begin{figure}[htbp]
{
\begin{center}
{
\scalebox{1}{ \input{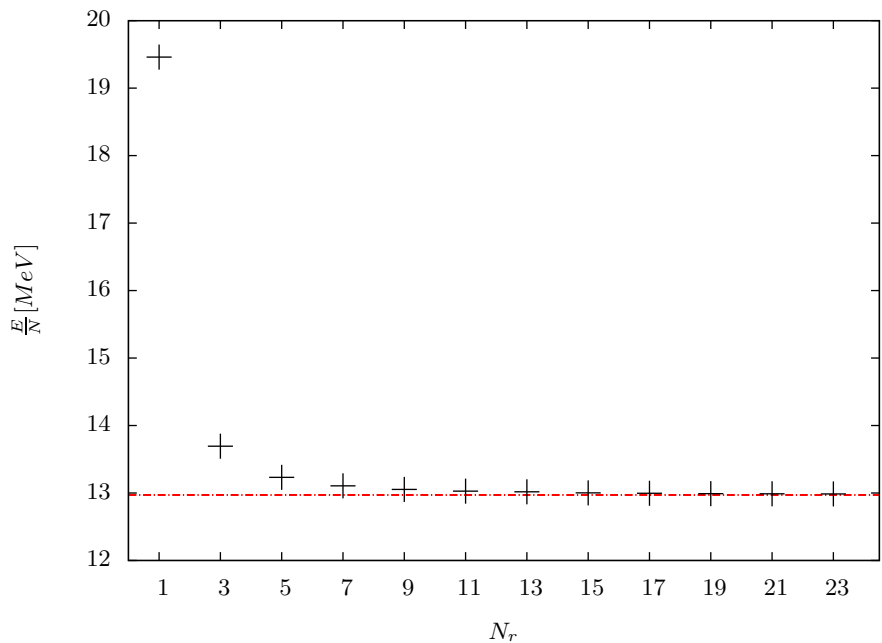}}
}		 
\end{center}
\caption{
Energy associated to a system of two free FMD particles  in one dimension at a density $\overline{\rho}=\frac{1.37}{\pi}$ as a function of the number of replicas. Dash-dotted line: free Fermi gas in one dimension.
}
\label{free_tak}
}
\end{figure}

The FMD ground state energy for a system of two free particles in one dimension is shown in  
Figure \ref{free_tak} as a function of the number of replicas, with the Wannier choice for the wave packets. For each calculation, the number $M$ was chosen high enough to have convergent results. 
We have already observed that the FMD variational space is sufficiently flexible to describe a free-particles system. The FMD solution for each value of $N_r$ thus represents the exact energy of a finite system with
periodicity $L=2/\overline{\rho}$ and size $L_{tot}=L\cdot N_r$. The thermodynamic limit, corresponding to the anlytical Fermi gas result, is obtained for about ten replicas, corresponding to ten values for the phase  $\gamma$ of eq.(\ref{gamma}).

\section{Comparison of the two methods}

In the previous sections, we have introduced the Bloch and the Wannier representation as two alternative ways to address the thermodynamical limit in variational theories. From a principle point of view the two approaches are very similar, however they lead to very different expressions in the calculation of observables.
In particular, the fact of considering only diagonal terms in the phase quantum number naturally emerges as a consequence of the orthogonality of the Wannier basis in the replica method, while it appears as a simple working hypothesis in TABC. Moreover, the equivalence is expected in the ideal case of an infinite number of twists and under the condition that the wave function is an exact eigenvalue of the Hamiltonian. This first condition is numerically expensive, while the second is not assured in variational methods.
For all these reasons, the equivalence of the two formalisms is not a-priori clear and will be discussed in the present section.

\subsection{Finite number of replicas and finite number of twists}

We now show that, in the hypothesis that the variational theory produces the exact eigenvectors of the many-body Hamiltonian, the two methods to treat the antisymmetrisation correlations can be mapped on each other for any system size.

In the Bloch formalism, if we impose that the condition of periodic boundaries eq.(\ref{periodic}) is verified 
for a finite translation of size $N_r L$, the original problem of an infinite system with periodicity $L$ is transformed into the problem of a finite system of size $N_r L$, with periodic boundary conditions.
 
\begin{figure}[htbp]
{
\begin{center}
{
\scalebox{1}{ \input{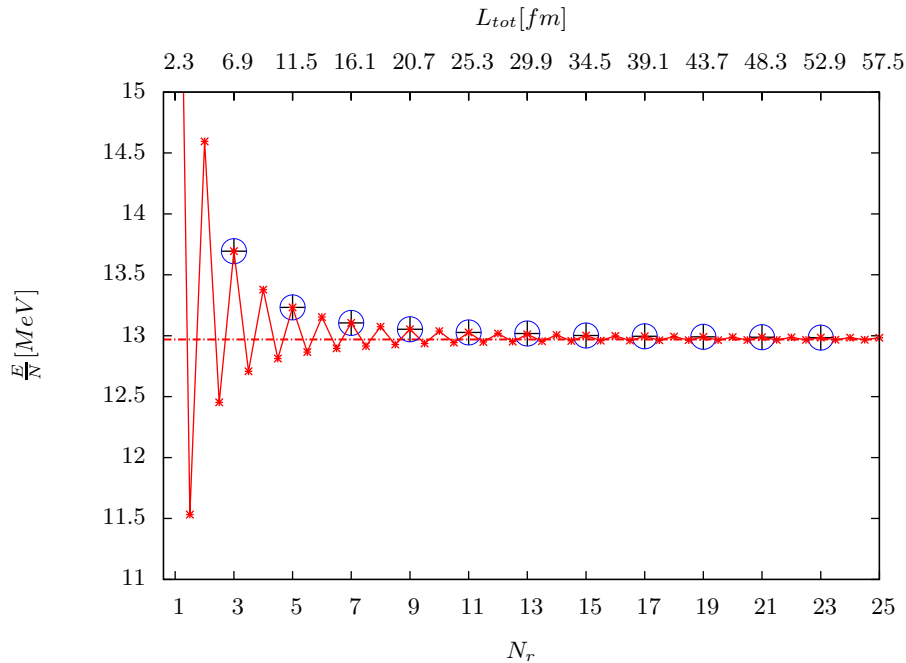}}
}		 
\end{center}
\caption{
Energy of a free  periodic   one-dimensional system at a density $\overline{\rho}=\frac{1.37}{\pi}$ as a function of the number of replicas. The upper abscissa represents the total linear size in the case of simple periodic boundary conditions (full line), the lower one  is a number of replicas for the replica method (crosses), and a number of twists for the Bloch method (circles). Dash-dotted line: free Fermi gas in one dimension. 
}
\label{free_all}
}
\end{figure}

The allowed values for $\theta$ are then discrete $\theta=\frac{2n-N_r}{N_r}\pi$, where $n$ is an integer.
This amounts to discretize the integral (\ref{TABC}) with  $\Delta \theta= \frac{2\pi}{N_r}$. 
We expect then the two methods to give exactly the same results.
This expectation is confirmed by Figure \ref{free_all}, which compares for the case of the non-interacting one-dimensional system the energy obtained considering a periodic system of 
$N_{tot}=\overline{\rho} N_r L$ particles, with the calculation of a reduced $N=\overline{\rho} L$ 
system averaged over $N_r$ phases with the two (Bloch and replica) proposed methods. We can see that the
agreement is perfect for all sizes, and all calculations converge to the free Fermi gas result with increasing number of replicas (respectively: twists).
 
This shows the equivalence of the different methods when the wave vector is an eigenvalue of the Hamiltonian.

\subsection{Limitations of the Bloch method in variational applications}

In realistic applications of interacting fermion system, the variational solution is always an approximation of the exact eigenvalue. This may create a bias in the different method that we turn to discuss. 
For eq.(\ref{transl_op}) to correctly account for the correlations due to the particles outside the cell $L$, it is necessary that the wave function $\Psi_\theta$ is an eigenfunction of the hamiltonian.
Conversely the Wannier representation can be obtained from the solution of the infinite periodic system via a simple linear transformation eq.(\ref{uu}). Of course if the variational ansatz is not adequate to describe the exact state both representations will be false. However the error due to the inadequacy of the solution in the cell $L$ will be propagated in an uncontrolled way in the Bloch method for the description of the global system, and this will not be the case for the replica method that directly addresses the replicated
system of size $N_r L$.

\begin{figure}[htbp]
{
\begin{center}
{
\scalebox{1}{ \input{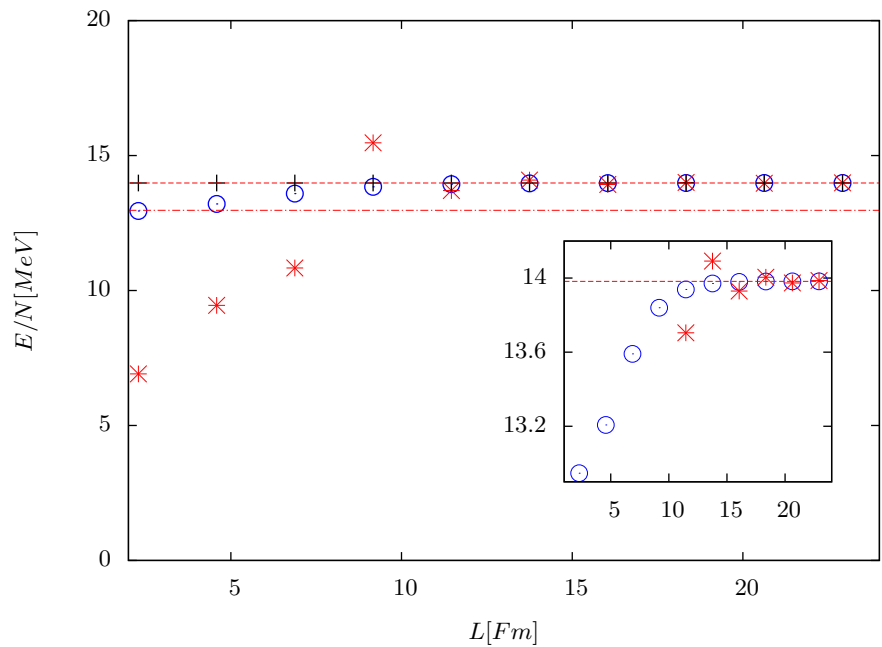}}
}		 
\end{center}
\caption{
Comparison of the different boundary conditions for the AMD model 
in the case of a non interacting system as a function of the system size.  
Dash-dotted line: free Fermi gas. Dashed line: AMD ground state energy for a system size large enough for finite size effects be negligible ($L=150$).
Circles: finite system with Bloch boundary conditions. Cross: replica method. Stars: periodic boundary conditions.}
\label{fmd-amd}
}
\end{figure}

To illustrate this difference, we can once again take the simple example of the non-interacting Fermi gas.
There is a  popular microscopic theory in nuclear physics, the so-called AMD model\cite{amd}, where the single particle wave packets are gaussians of fixed width. With this ansatz the variational space does not contain the exact solution of this problem at the bulk limit, that is an antisymmetrized combination of plane waves. 
In particular, if we choose a small value for the width, as it is variationally obtained to describe finite nuclei\cite{amd}, it is clear that the states will be very far from the exact eigenvectors of the kinetic hamiltonian.
The comparison of the two methods in this model case is shown in Figure \ref{fmd-amd}.

We can see in Figure\ref{fmd-amd} that the AMD ansatz is not adapted for this problem, from the fact that asymptotically the energy of the free Fermi gas (full line) is not recovered by the AMD calculation (dashed line). 
This asymptotic value is perfectly reproduced by the replica method (crosses) for any system size, while the Bloch method (open circles) produces artificial fluctuations and attains convergence only when
the Bloch phase is negligibly small, and the calculation is perfectly equivalent to simple periodic boundary conditions (stars).

The non-interacting system is certainly not the ideal application ground of molecular dynamics model.
As a second model case which is closer to the physical case of the neutron star crust we take a one-body periodic potential, which is obtained as a periodic self-consistent mean field in realistic applications of mean-field variational models \cite{San04,Kha08,Gra08,Mon07,douchin} if the two-body nuclear and Coulomb interaction is added, including the interaction with a uniform electron background. For this model one-dimensional application we only wish to discuss the effect of the boundary conditions, therefore we restrict to the simpler case of an external periodic potential. In particular, the choice of a gaussian potential has the advantage of making all calculations analytical. Let us take a potential form:
\be
V_{gaus}(\hat{x})=V_0 \sum_{n=-\frac{N}{2}}^{\frac{N}{2}}\sum_{m=-\infty}^{\infty}
 exp \left[
  -\frac{1}{2a} \left( \hat{x}-b_{n}-mL  \right)^2  
  \right] 
\ee
where $V_0,a,b_n$ are parameters. In particular, the choice $a=\frac{4}{\overline{\rho}}$, $b_{n}=\frac{n}{\overline{\rho}}$ for a system with average density  $\overline{\rho}$, corresponds 
to having a single particle per potential well.   
The average energy is readily calculated as
\begin{eqnarray}
\left< \hat{V}_{gaus} \right>_{\Psi} 
&=&
V_0
\lim_{N_r\rightarrow \infty}\frac{1}{N_r}
\sum_{i,j=1}^{N} B^{-1}_{ij}
\sqrt{2\pi \frac{a_ia_j^*a}{a_j^*a+a_ia+a_ia_j^*}} \nonumber
 \\
&&
\sum_{m_1,m_2,n=-\infty}^{\infty}
exp \left[ -
\frac{a_j^*(Z_i-b_{n}+m_1L)^2 + a_i(Z_j^*-b_{n}+m_2L)^2 + a(Z_i-Z_j^*-(m_1-m_2)L)^2}
{a_j^*a+a_i a+a_ia_j^*}
    \right]   
\end{eqnarray}

\begin{figure}[htbp]
{
\begin{center}
{
\scalebox{1}{ \input{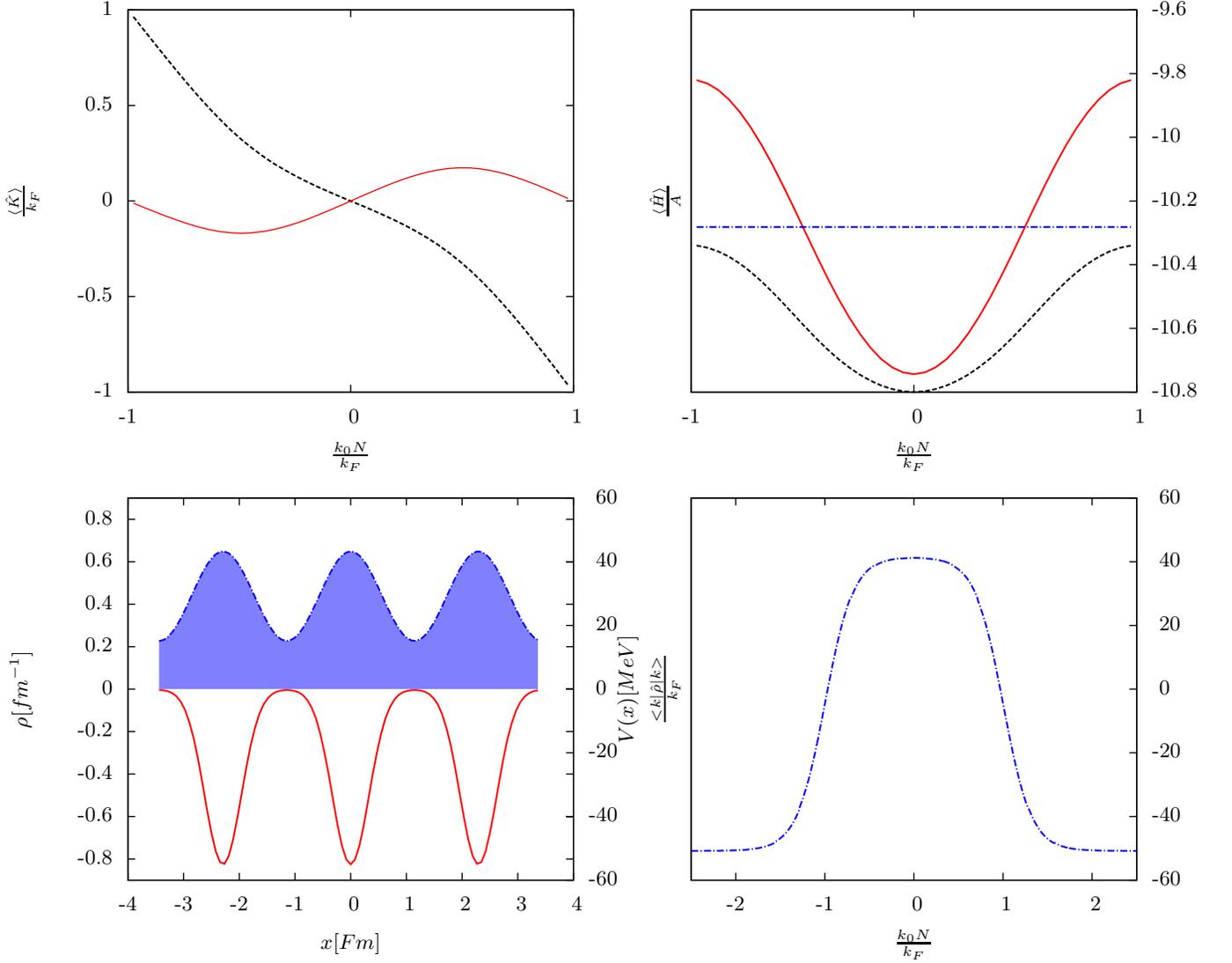}}
}		 
\end{center}
\caption{
FMD calculation of a system of three particles in a one dimensional box of size $L$ subject to a periodic external potential  with Bloch boundary conditions. Upper part:
behavior of the total linear momentum (left) and total energy (right)  as a function of the Bloch wave number $k_0=\theta/L$ 
Full lines: total expectation values $\langle\hat K\rangle_\Psi,\langle\hat H\rangle_\Psi$;
dash lines: averages over the variational part of the wave function  $\langle\hat K\rangle_\Phi,\langle\hat H\rangle_\Phi$ without the contribution of the Bloch phase. Dashed-dotted lines: expectation values including the phase average.
Lower part:
dash-dotted lines: spatial (left) and momentum (right) density;
full line: gaussian potential.
}
\label{pot}
}
\end{figure}

Results from the Bloch method for a case of 3 particles with an average density $\overline{\rho}=\frac{1.37}{\pi}$ are presented in Figure \ref{pot}.  

At variance with the free particle case, the periodic part of the wave function $\Phi$ now explicitly depend on the $\theta$ quantum number, leading to a $\theta$ dependence of the one body density.

\begin{figure}[htbp]
{
\begin{center}
{
\scalebox{1}{ \input{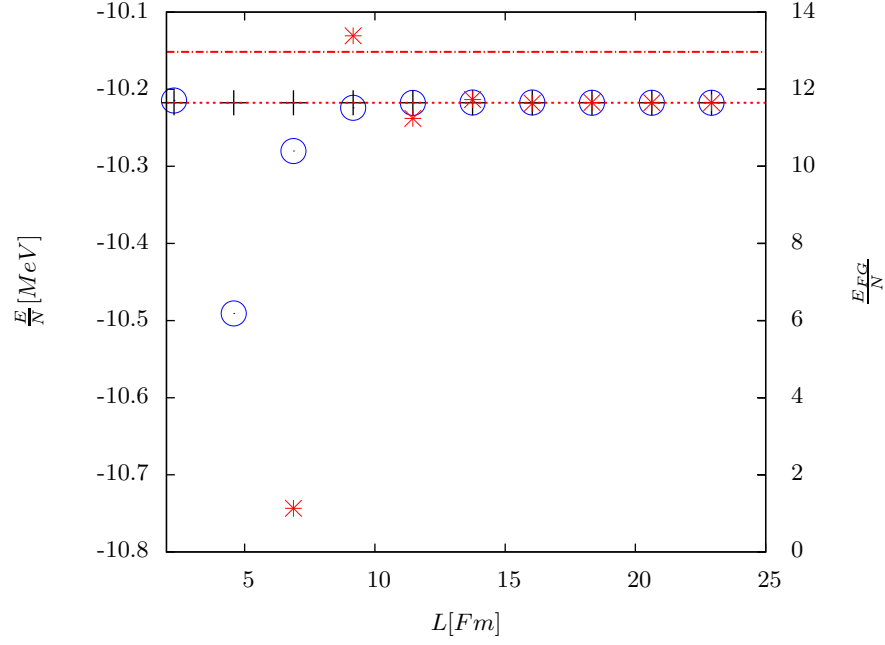}}
}		 
\end{center}
\caption{Comparison of the different boundary conditions for the FMD model 
in the case of a system subject to an external gaussian potential as a function of the system size.  
Dash-dotted line: free Fermi gas. Dashed line: FMD ground state energy for a system size large enough for finite size effects be negligible ($L=30$). 
Circles: finite system with Bloch boundary conditions. Cross: replica method. Stars: periodic boundary conditions.
 }
\label{epot}
}
\end{figure}

The comparison between the different methods of treating the boundary conditions is shown in figure \ref{epot}. 

This model is less trivial than the free Fermi gas, although still very schematic. The FMD model is expected to be a good approximation of this system, but there is no guarantee the FMD solution should be exact.
Similar to the previous application, the replica method gives by construction the asymptotic result for any finite size, provided a sufficient number of phases is considered. Conversely for the Bloch method, which is calculated with the same number of phases 
as the replica method for the application of Fig.\ref{epot}, a convergence is reached only when the effect of the phase can be neglected.

This model example shows that in realistic applications, where any variational ansatz might be simply an approximation of the exact energetics of the system, the different boundary conditions are not equivalent, and the employ of Bloch corrections to the single particle basis may increase the deviation respect to the exact solution.

\section{Conclusions}

In this paper we have introduced and critically discussed two different ways of addressing the boundary conditions in finite fermionic system in order to account for the infinite range antisymmetrisation correlations present at the thermodynamical limit. 
In the case of state vectors  which are eigenstates of the many-body Hamiltonian, we have shown that the use of Bloch wave functions with twist averaged boundary conditions physically corresponds exactly to a bigger system constituted of a number of replicas of the system under study equal to the number of phases. In turn, this last formalism can be interpreted as the use of localized Wannier states. While these different representations are equivalent when dealing with eigenvectors, the same is not true 
in the case of approximate solutions of the many-body problem, as it is the case in the variational approach. 
In this case, the approach of the thermodynamic limit is not properly treated by the Bloch method, while the replica method provides a very accurate evaluation of the Fock energy. Even in the case of a non-orthogonal single particle basis this method can be numerically implemented with a moderate computational cost, opening the possibility of a completely quantum-mechanical treatment of nuclear matter with molecular dynamics approaches.

\begin{acknowledgments}
{This paper has been partly supported by ANR under the project NEXEN.

Discussions with Klaas Vanternhout and Karim Hasnaoui are gratefully acknowledged.}
\end{acknowledgments}


\end{document}